\documentclass[aps,prb,twocolumn]{revtex4-1}
\usepackage{amssymb}
\usepackage{graphicx}
\usepackage{amsmath}
\usepackage{amssymb}
\usepackage{amsfonts}
\usepackage{graphicx}
\usepackage{epsfig}
\usepackage{hyperref}
\usepackage{color}
\usepackage{xcolor}
\usepackage{epstopdf}
\usepackage{siunitx}
\hypersetup{
colorlinks=true,
urlcolor=blue,
citecolor=blue,
linkcolor=blue,}

\begin{document}

 \title{A de Hass-van Alphen study of the Type-II Dirac semimetal candidates $A$Te$_2$ ($A =$ Pt, Pd)}

 \author{Amit$^1$, R. K. Singh$^1$, Neha Wadhera$^2$, S. Chakraverty$^2$, and Yogesh Singh$^1$}
 \affiliation{$^1$ Department of Physical Sciences,
 	Indian Institute of Science Education and Research Mohali,
 	Sector 81, S. A. S. Nagar, Manauli, PO: 140306, India.\\
	$^2$ Nanoscale Physics and Laboratory, Institute of Nanoscience and Technology, SAS Nagar, Mohali, Punjab 140062, India.}

 \date{\today}

\begin{abstract}
We report on a magneto-transport and quantum oscillations study on high quality single crystals of the transition metal di-tellurides PtTe$_2$ and PdTe$_2$.  The de Haas-van Alphen (dHvA) oscillations in the magnetization measurements on PtTe$_2$ reveal a complicated, anisotropic band structure characterized by low effective masses and high mobilities  for the carriers.  Extracted transport parameters for PtTe$_2$ reveal a strong anisotropy which could be related to the tilted nature of Dirac cone. Using a Landau level fan diagram analysis we find at least one Fermi surface orbit with a Berry phase of $\pi$ consistent with Dirac electrons for both PtTe$_2$ and PdTe$_2$.  The light effective mass and high mobility are also consistent with Dirac electrons in PtTe$_2$.  Our results suggest that similar to PdTe$_2$, PtTe$_2$ might also be a three dimensional Dirac semimetal.

\end{abstract}
\maketitle

Recent discovery of topological semimetals (TSM) in three dimensions, also known as 3D version of graphene, has intensified the search for this unique state of matter in various magnetic and non-magnetic stoichiometric materials \cite{1, Ando,2,3, 4}. Graphene is considered topologically trivial due to an even number of band crossings at the Fermi level.  Graphene can be gapped or localized by disorder. On the other hand Dirac/Weyl semimetals (DSM/WSM) belong to the topological class of metals having an odd number of bulk bands with a linear dispersion in all three momentum directions and are protected from gapping by certain symmetries\cite{4, 5,6,7,8,9,10,11}. A DSM can be transformed in to a WSM by breaking time reversal symmetry, inversion symmetry or both. Near the bulk Dirac point the low energy excitations mimic the relativistic massless quasi-particles predicted theoretically in the context of high energy physics. TSMs have been shown to exhibit exotic physical properties like the chiral anomaly, non Ohmic transport, nonlocal conduction, a Berry phase of $\pi$, and many other anomalous optical and transport phenomena. Two forms of TSMs have been predicted theoretically, depending on the nature of the bulk Dirac dispersion relation. They are categorized  as Type-I and Type-II TSMs.  While Type-I has linear and isotropic dispersion in the momentum space, the dispersion relation in the Type-II topological semimetals is tilted. 
 
The linear Dirac dispersion in the bulk band structure can be probed by studying quantum oscillations in transport experiments as has been demonstrated for graphene, topological insulators and recently in the Dirac and Weyl semimetals \cite{Ando}. Previous quantum oscillation studies on Weyl and Dirac semimetals have demonstrated a connection between band topology and the phase acquired by the charge carriers. The wave function of a relativistic quasi-particle acquires a nontrivial geometric phase of $\pi$ along the cyclotron orbits in a magnetic field, (also known as the $\pi$ Berry phase), which can be calculated from the quantum oscillations in resistivity and magnetization \cite{Ando}. 
 
The tilting of the Dirac cone has been predicted in PtTe$_2$ and PdTe$_2$ as well \cite{12,13,14,15,16,17,18,19}.  PdTe$_2$ has previously been confirmed to be a Dirac semimetal by angle resolved photoemission and quantum oscillation experiments \cite{14,16,17,20,21,22,23}.  In the transition metal di-telluride family experimental work focusing on demonstrating the Type-II DSM nature has been mainly on PdTe$_2$, PtSe$_2$, and Pd/PtSeTe compounds \cite{13,18,20,24}.  However experimental evidence for the DSM nature of another compound from the same family PtTe$_2$ is missing. There are some photoemission spectroscopy reports regarding tilted nature of the Dirac cone in PtTe$_2$, but quantum oscillation studies on this compound are still missing, which can help establish the three dimensional Dirac character of this compound. The key signatures of Dirac carriers, such as low (ideally massless) effective mass of carriers, high mobility, and a Berry phase of $\pi$ can be extracted from quantum oscillation measurements\cite{9,10,11,25,26,27,28,29,30}.  
 
 In this article we have measured dHvA quantum oscillations on high quality single crystals of PtTe$_2$ and PdTe$_2$ for both out of plane (B//c) and in-plane (B//ab) configurations.  The temperature dependent resistivity measurements reveal a typical metallic character with high residual resistivity ratio (RRR) for the two cases.  The RRR for PdTe$_2$ is $\approx 238$ and that for PtTe$_2$ is $\approx 96$.   Both PtTe$_2$ and PdTe$_2$ show pronounced magnetization quantum oscillations with multiple frequencies which suggests multiple Fermi pockets crossing the Fermi level.  The number and positions of the frequencies suggest different band structures for PdTe$_2$ and PtTe$_2$.   For the orbit with the largest amplitude, the extracted Berry phase is close to the value $\pi$ for both PtTe$_2$ and PdTe$_2$.  The analysis of the observed quantum oscillations in both in plane and out of plane field orientations enables us to calculate important band parameters for the two semimetals. Low effective masses and high mobilities are estimated for both materials. Additionally, a Berry phase close to $\pi$ for at least one orbit strongly suggests the Dirac nature of bulk carriers in these two Dirac semimetals.

\subsection{Experimental}
\begin{figure}[t]
  	\includegraphics[width=3 in]{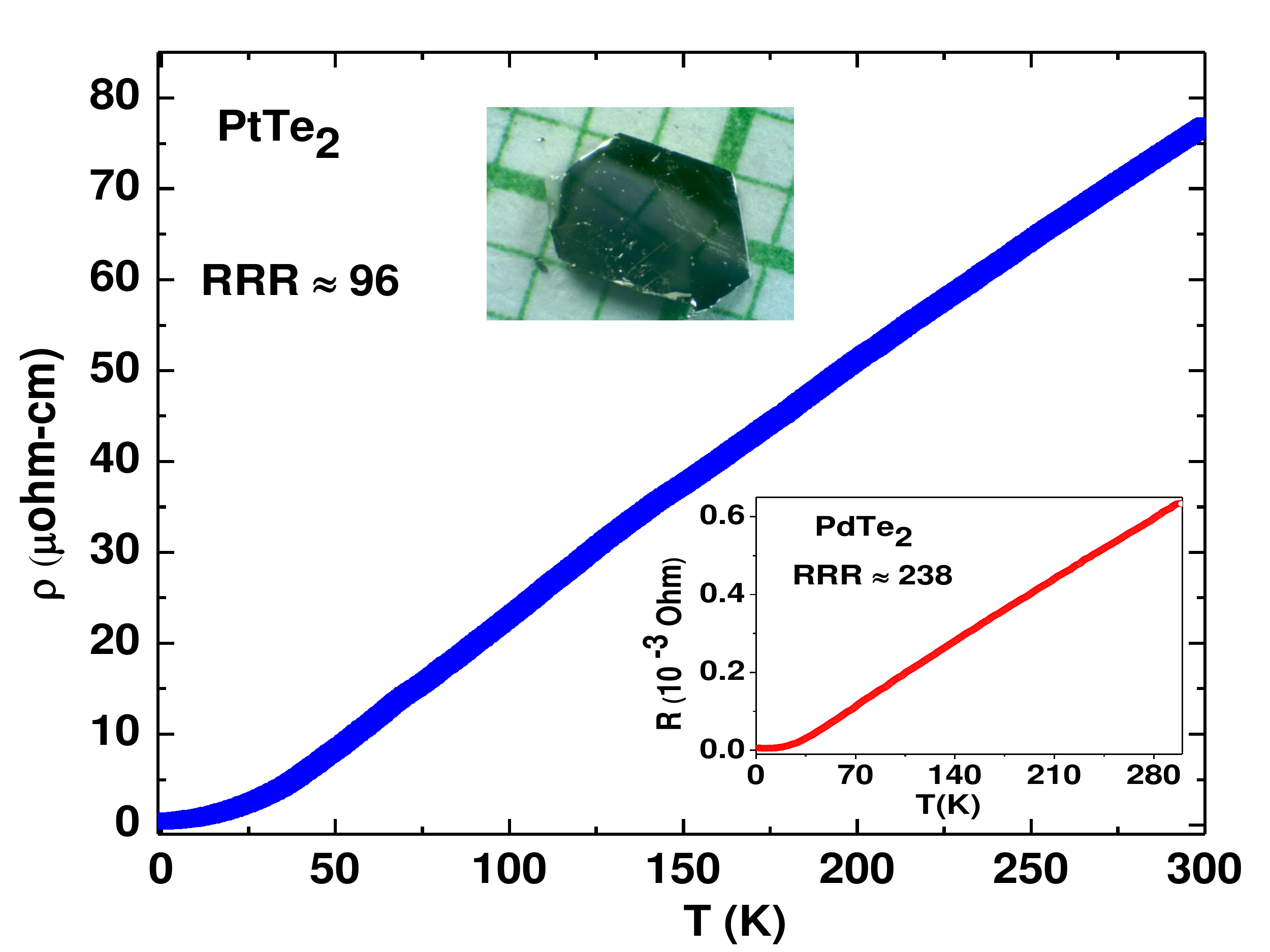}
 	\caption{ Temperature dependent electrical resistivity of  PtTe$_2$ and PdTe$_2$ (lower inset)  single crystals. A current I = 2 mA is applied in the crystallographic ab plane. Upper inset shows an optical image of a  PtTe$_2$ crystal placed on a millimeter grid.} 
\label{Fig-res}
 \end{figure}

High quality single crystals of PdTe$_2$ were synthesized by a modified Bridgman technique as reported earlier\cite{31}. For PtTe$_2$ crystals, the starting elements Pt powder (99.9 \%, Alfa Aesar) and Te lump (99.9999\%, Alfa Aesar) were taken in the molar ratio $2 : 98$ and sealed in an evacuated quartz tube. The tube was heated to \SI{790}{\celsius} in $15$~h, kept at this temperature for $48$~h in order to homogenize the solution and then slowly cooled to \SI{500}{\celsius} at a rate of \SI{2.5}{\celsius}/hr. The excess Te liquid was decanted isothermally for $2$~days.  Finally, the tube was allowed to cool down to room temperature by shutting off the furnace.  Plate-like hexagonal shaped crystals were obtained after breaking open the tube at room temperature.  A typical crystal of PtTe$_2$ is shown on a millimetre grid in the inset of Fig.~\ref{Fig-res}. The powder X-ray diffraction pattern obtained on crushed crystals of PdTe$_2$ and PtTe$_2$ confirms the phase purity and CdI2 type crystal structure with P3m1 (164) space group for both materials. The chemical composition and uniformity of stoichiometry for crystals of both compounds was confirmed by energy dispersive spectroscopy at several spots on the crystals used for the measurements reported in this work. The electrical transport and magnetic measurements (0-14 T) were performed on a physical property measurement system by Quantum Design (QD-PPMS).

\begin{figure}[t]
\includegraphics[width= 3.5 in]{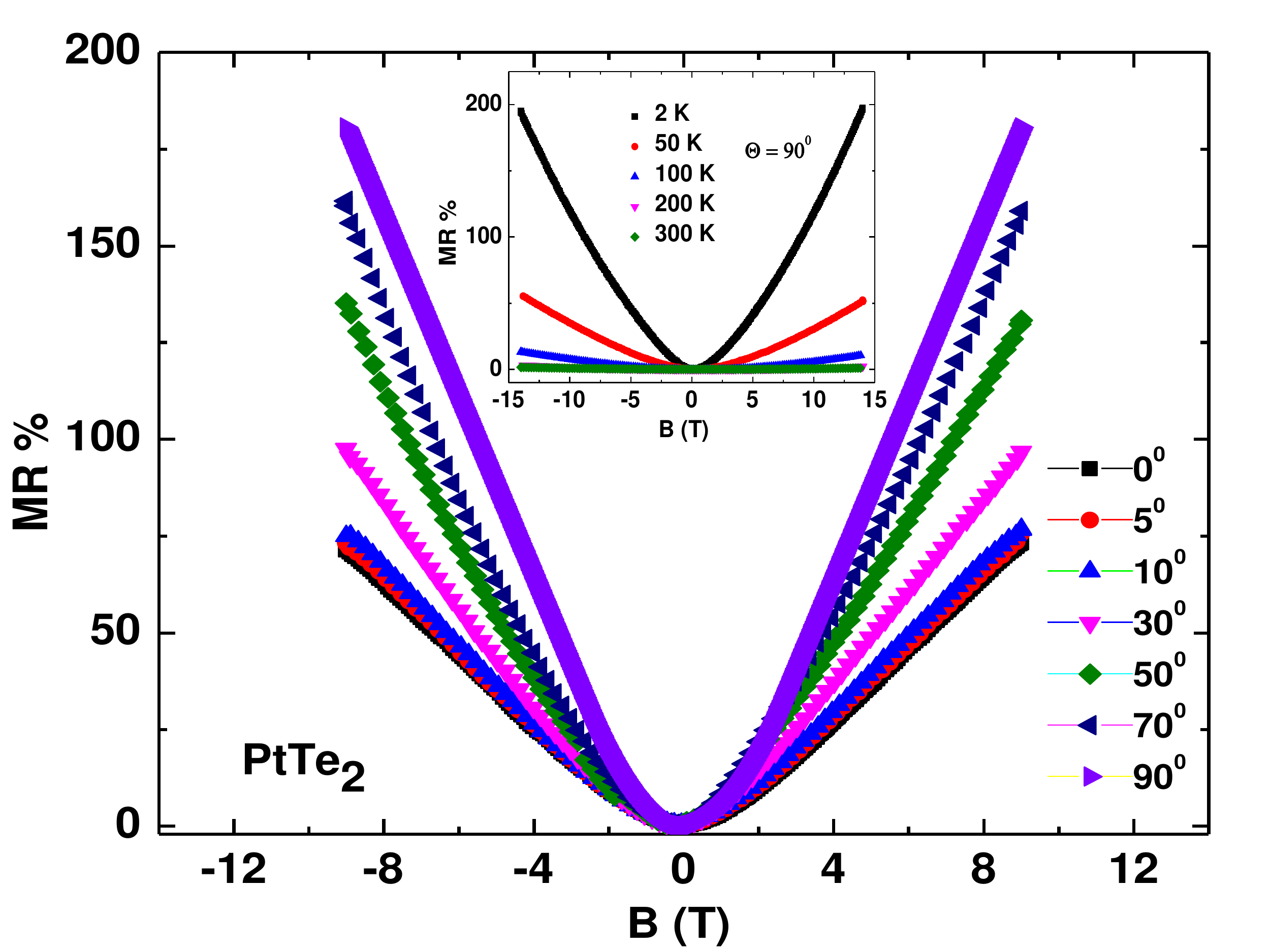}
\caption{ Magnetoresistance (MR) for PtTe$_2$ as a function of the magnetic field $B\leq 9$~T applied at various angles to the direction of the current $I$ which was always applied within the $ab$-plane.  Inset shows the MR measured at various $T$ with $B \perp I$ for $B \leq 14$~T\@.} 
\label{Fig-MR-Pt}
 \end{figure}
 
\subsection{Results}
\subsubsection{\rm PtTe$_2$}

 \begin{figure*}[t]
 	\includegraphics[width=1\textwidth]{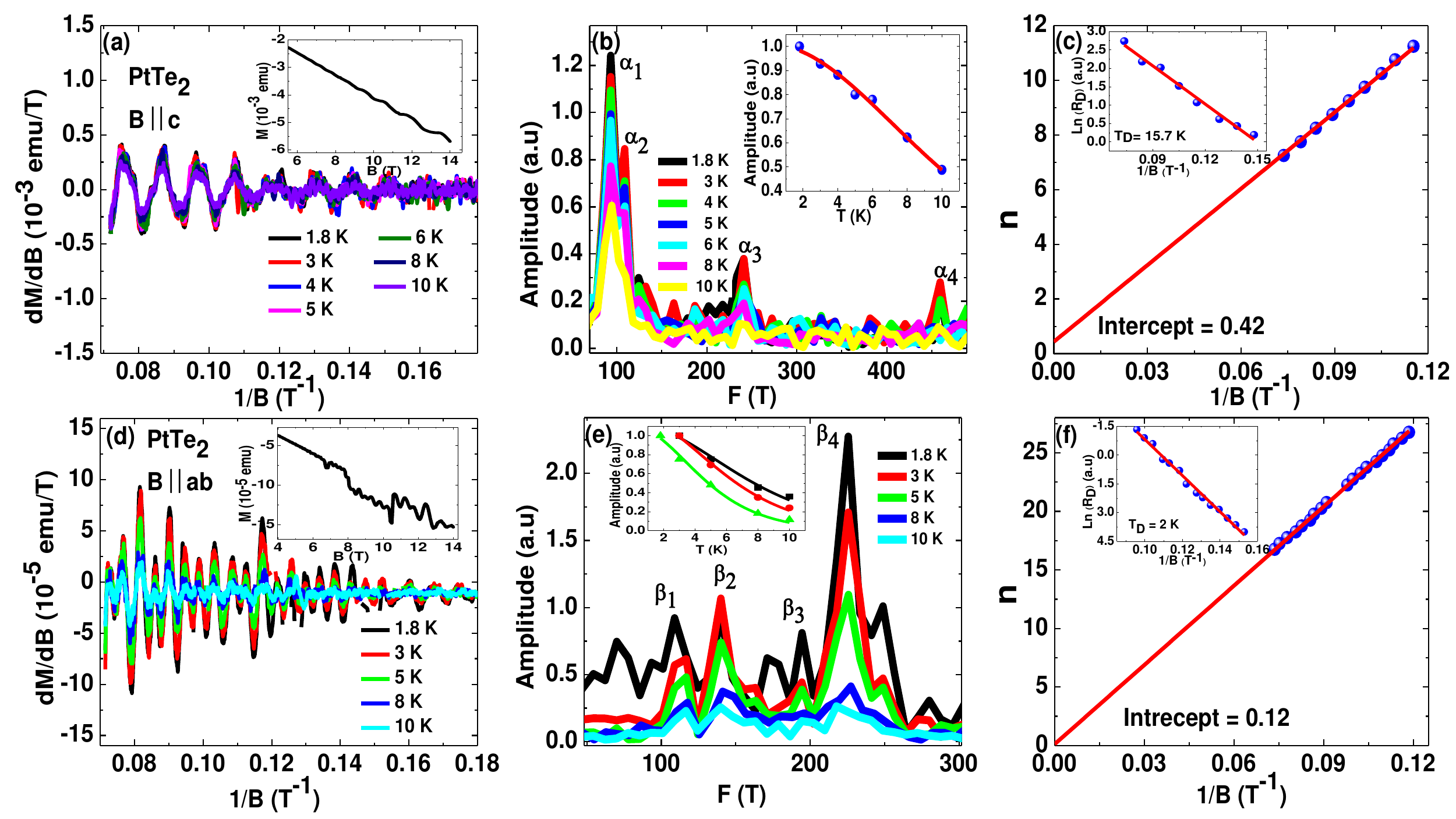}
 	\caption{(a and d)Isothermal magnetization oscillation data for PtTe$_2$ in the two directions B//c and B//ab, at various temperatures.Inset shows row magnetization data at 2K. (b and e) shows the temperature dependence of FFT spectra and fitting of the temperature dependent amplitude is shown in the inset for the two cases. (c and f) The landau level fan diagrams for the two cases. Inset shows the Dingle fitting. } 
 	\label{Fig-dM}
 \end{figure*}
 
Figure~\ref{Fig-res} shows electrical resistivity $\rho$ versus temperature $T$ for a single crystal of PtTe$_2$ measured with an ac current of amplitude $I = 2$~mA applied within the $ab$-plane of the crystal.  Metallic behaviour is observed in the whole $T$ range ($1.8$~K to $300$~K) of measurements.  The residual resistivity ratio RRR $= \rho(300~K)/\rho(1.8~K) \approx 96$ indicates the high quality of the crystal.  The top inset shows an optical image of an as-grown PtTe$_2$ crystal showing the hexagonal morphology of the underlying crystal structure.  The bottom inset shows the electrical resistance $R$ vs $T$ for a crystal of PdTe$_2$.  The RRR for this PdTe$_2$ crystal was $\approx 238$.  A high RRR has been regarded as an indication of relativistic charge carriers in three dimensional Dirac semimetals such as Cd$_3$As$_2$, NbAs, PdTe$_2$, NbP etc \cite{16,25,26,27,30,31,32,33,34}. 

Figure~\ref{Fig-MR-Pt} shows the magneto resistance (MR) data on the PtTe$_2$ crystal.  The main panel shows the magnetic field $B$ dependence of the MR measured at $T = 1.8$~K as a function of the angle between $B \leq 9$~T and the direction of the electrical current $I$, which was always applied in the same direction within the $ab$-plane of the crystal.  The MR for all angles increases monotonically and tends to a linear in $B$ behaviour.  The magnitude of MR reaches about $200\%$ which is smaller than observed for other DSMs.  We specifically point out that no negative contribution to the MR was observed for $B \parallel I$ indicating the absence of the Chiral anomaly.  The MR as a function of $B \perp I$ at various temperatures $T$ is shown in the inset of Figure~\ref{Fig-MR-Pt} and shows that the magnitude of MR has a strong $T$ dependence.

The isothermal magnetization $M$ data for PtTe$_2$ plotted as $dM/dB$ versus $1/B$ at different temperatures for magnetic fields $B \leq 14$~T applied along the $c$-axis ($B \parallel c$) and applied within the $ab$-plane ($B \parallel ab$) are shown in Figs.~\ref{Fig-dM}~(a) and (d), respectively.  The raw $M$ vs $B$ data for the two field directions are shown in the insets.  The magnetization data reveal pronounced dHvA oscillations starting from $4$~T. The low onset field value of the quantum oscillations also point to the high quality of the PtTe$_2$ crystal.  Pronounced periodic oscillations as a function of $1/B$ are clearly visible up to $10$~K in both field orientations.  Multiple frequencies for both field orientations were observed in the fast Fourier transform (FFT) spectra of the quantum oscillations as shown in Figs.~\ref{Fig-MR-Pt}~(b) and (e). The multiple frequencies in the FFT spectra indicates the presence of multiple Fermi surface pockets at the Fermi level. Additionally, the presence of dHvA oscillations for both $B \parallel c$ and $B \parallel ab$ directions, confirms the presence of a three dimensional Fermi surface in PtTe$_2$. From the temperature dependent FFT spectra, we find four main frequencies for both $B \parallel c$ and $B \parallel ab$ as shown in Figs.~\ref{Fig-MR-Pt}~(b) and (e). The main frequencies for the $B \parallel c$ are labeled as $\alpha_1 (93.3$~T), $\alpha_2 (108.9$~T), $\alpha_3 (241$~T) and $\alpha_4 (459$~T).  The main frequencies for $B \parallel ab$ are labelled $\beta_1 (116.6$~T), $\beta_2 (140.6$~T), $\beta_3 (194.5$~T) and $\beta_4 (225.6$~T). The information regarding the Fermi surface area corresponding to these frequencies can be determined by the Onsager relation $F =A_{F} (\varphi/2\pi^2 )$, where $\varphi = h/e$, is the magnetic flux quantum and $A_{F}$ is the Fermi surface area. The calculated Fermi surface area for the frequencies extracted for the two field orientations are listed in Table~\ref{Table 1}. 
 
A quantitative analysis of the dHvA oscillations can be made using the Lifshitz-Kosevich (LK) equation which gives the oscillatory contribution to the magnetization as:

\begin{equation}
\Delta M \propto - R_{T}R_{D} {\rm sin}(2\pi[{F\over B}-({1\over 2}-\phi)]) 
\end{equation}

where, $R_{D} = exp(-\lambda T_{D})$ is the Dingle factor, $T_{D} = \hbar/2 \pi K_{B} \tau$ is the Dingle temperature, and the temperature dependent damping of the oscillations is accounted for by the factor $R_{T} = \lambda T/{\rm sinh}(\lambda T)$,  with $\lambda = (2\pi^2 K_{B} m^*/\hbar eB)$ and $m^*$ the effective cyclotron mass.  The phase $\phi = \phi_B/2\pi -\delta$, where $\phi_B$ is the Berry phase and $\delta$ is an extra phase factor.  The value of this additional phase shift $\delta$ depends on the dimensionality of the Fermi surface and takes the value $0$ or $\pm 1/8$ ($-$ for electron like and $+$€œ for the hole like) for two and three dimension, respectively \cite{26,27,28,36}.  

The damping factors $R_{T}$ and $R_{D}$ can be used to calculate important band parameters such $m^*$, life time $\tau$ and mobility $\mu$ of the carriers. The values of $\tau$ (corresponding to Dingle temperature $T_{D}$ ) extracted from fitting of the $B$ dependence (see inset of Figs.~\ref{Fig-MR-Pt}~(c) and (f)) of the oscillation amplitude are given in the Table~\ref{Table 2}. For $B\parallel c$ the value of $\tau = 7.7\times10^{-14}$~s, and for $B\parallel ab,~ \tau = 6.0\times10^{-13}$~s.  The corresponding mobility $\mu = e\tau/m^*$ in the two directions are estimated to be $902.0~cm^2/$Vs and $3296~cm^2$/Vs, respectively.  These values of $\mu$ are comparable to values reported previously for some Dirac semimetals \cite{20,27,28,35,36}. The large difference in the mobility (more than a factor of three) in the two crystal orientations suggests that the dynamics of carriers in PtTe$_2$ is highly anisotropic with carriers in the $c$-axis direction being less mobile than in the $ab$-plane. This would be consistent with a tilted Dirac cone as a tilting of the cone in one momentum direction might result in strong anisotropy in the transport properties as observed above for PtTe$_2$. 
 
The value of the effective mass for the two field orientations is found from the fitting of the temperature dependence of the amplitudes of the frequencies identified in the FFT spectra of the dHvA oscillations.  The fitting shown in the insets of Figs.~\ref{Fig-dM}~(b) and (e) was successful only for certain frequencies as the amplitude dropped too rapidly with $T$ for some frequencies.  The effective masses thus obtained are listed in the Table~\ref{Table 1}. For $B\parallel c$ the value of the effective mass $m^*$ corresponding to the frequency $\alpha_1 =93.3$~T is $0.15$.  The low value of $m^*$ for this frequency suggests the presence of relativistic charge carriers, which can further be confirmed by the Landau level fan diagram analysis which we present later. The values of $m^*$ corresponding to frequencies in the $B\parallel ab$ orientation are relatively larger in magnitude as listed in Table~\ref{Table 1}. 

	\begin{table}	
\caption{Fermi surface parameters for PtTe$_2$ obtained from the dHvA data shown in Figs.~\ref{Fig-dM}~(b) and (e)}.
		\scalebox{0.84}{
			\begin{tabular}{|c|c|c|c|c|c|} 
				\hline
				Compound &	F (T) & $A_f (\AA^{-2})(10^{-2})$ & $K_f (\AA^{-1}) (10^{-2})$ & $m^*/m$ & $V_f (m/s) (10^{5})$\\ [0.ex] 
				\hline
				$PtTe_2, 
				B \parallel c$&	93.3 & $0.89$ &  $5.3 $& 0.15& $4.1$\\ 
				\hline
				&108.9 & $1$  &$5.7$&$-$& $-$\\
				\hline
				&241 & $2.3$  &$8.6$&$-$& $-$\\
				\hline
				&459 & $4.4$  &$11.8$&$-$& $-$\\ 
				\hline
				$PtTe_2, 
				B \perp c$&	116.6 & $1.1$ &  $5.96 $& 0.21& $3.27$\\ 
				\hline
				&140.6 & $1.3$  &$6.54$&0.26& $2.9$\\
				\hline
				&194.5 & $1.9$  &$7.7$&$-$& $-$\\
				\hline
				&225.6 & $2.1$  &$8.3$&0.32& $3$\\[1ex] 	\hline
				
			\end{tabular}}
			\label{Table 1}
			\end{table}				
				
				\begin{table}	
				\caption{Parameters obtained from a Dingle fitting of the dHvA data shown in Fig.~\ref{Fig-dM}}.			
				\begin{tabular}{|c|c|c|c|c|} 
					\hline
					Compound &	$T_D (K)$ & $\tau (10^{-13} s)$ &  l (nm) & $\mu (cm^2/V-s)$\\ [0.ex] 
					\hline
					$PtTe_2,  B \parallel c$&	15.7 & $0.77$ &  31.57& 902\\ 
					\hline
					
					$PtTe_2, 	B \perp c$&	2 & $6$ &  180& 3296\\ 	[1ex] 	\hline

				\end{tabular}
				\label{Table 2}
				\end{table}					
				
We now present a Landau fan diagram analysis to estimate the Berry phase.  The presence of multiple frequencies (from multiple Fermi surface orbits) in the quantum oscillations makes it difficult to isolate contributions to the Berry phase from individual orbits.  We therefore take in to account the frequencies with the largest FFT amplitudes ($\alpha_1 = 93.3$~T for the $B\parallel c$ and $\beta_4 = 225.6$~T in case of $B\parallel ab$) as shown in the Figs.~\ref{Fig-dM} ~(b) and (e).  To construct the Landau level fan diagram we assign the landau index $n-1/4$ to the minima of quantum oscillations.  We are able to reach the $7^{\rm th}$ landau level in the $B\parallel c$ and $16^{\rm th}$ landau level for the $B\parallel ab$ configuration as shown in the Figs.~\ref{Fig-dM}~(c) and (f). The extrapolated value of the intercept ($= \phi_B/2\pi \pm \delta$) on the $n$ axis is found to be $0.42(2)$ and $0.12(1)$ for $B\parallel c$ and $B\parallel ab$, respectively.   The slopes obtained from the fits are $94.7$~T and $228.2$~T for the two directions.  These slopes are very close to the frequencies of the $\alpha_1$ and $\beta_4$ orbits, proving that we are analysing these orbits in the Landau level fan diagram \cite{Ando}.  From the value of the intercepts found above the estimated Berry phase for $B\parallel c$ and $B\parallel ab$ directions are $1.08(4)\pi~ [{\rm or}~ 0.59(4)\pi$] and $0.49(2)\pi~ [{\rm or}~ -0.01(2)\pi$], respectively. The value $\phi_B = 1.08(4)\pi$ for the $\alpha_1$ orbit in $B\parallel c$ direction is very close to the value $\pi$ expected for Dirac electrons.  Whereas, the value $\phi_B = 0$ for $B\parallel ab$ suggests that the orbit $\beta_4$ is trivial. A possible reason for the trivial character of bands in the $B\parallel ab$ direction can be the nonlinear nature of the bands owing to the Fermi level being away from the Dirac point. 
 
\subsubsection{\emph {PdTe}$_2$} 
\begin{figure}[t!]
 	\centering
 	\includegraphics[width=0.52\textwidth]{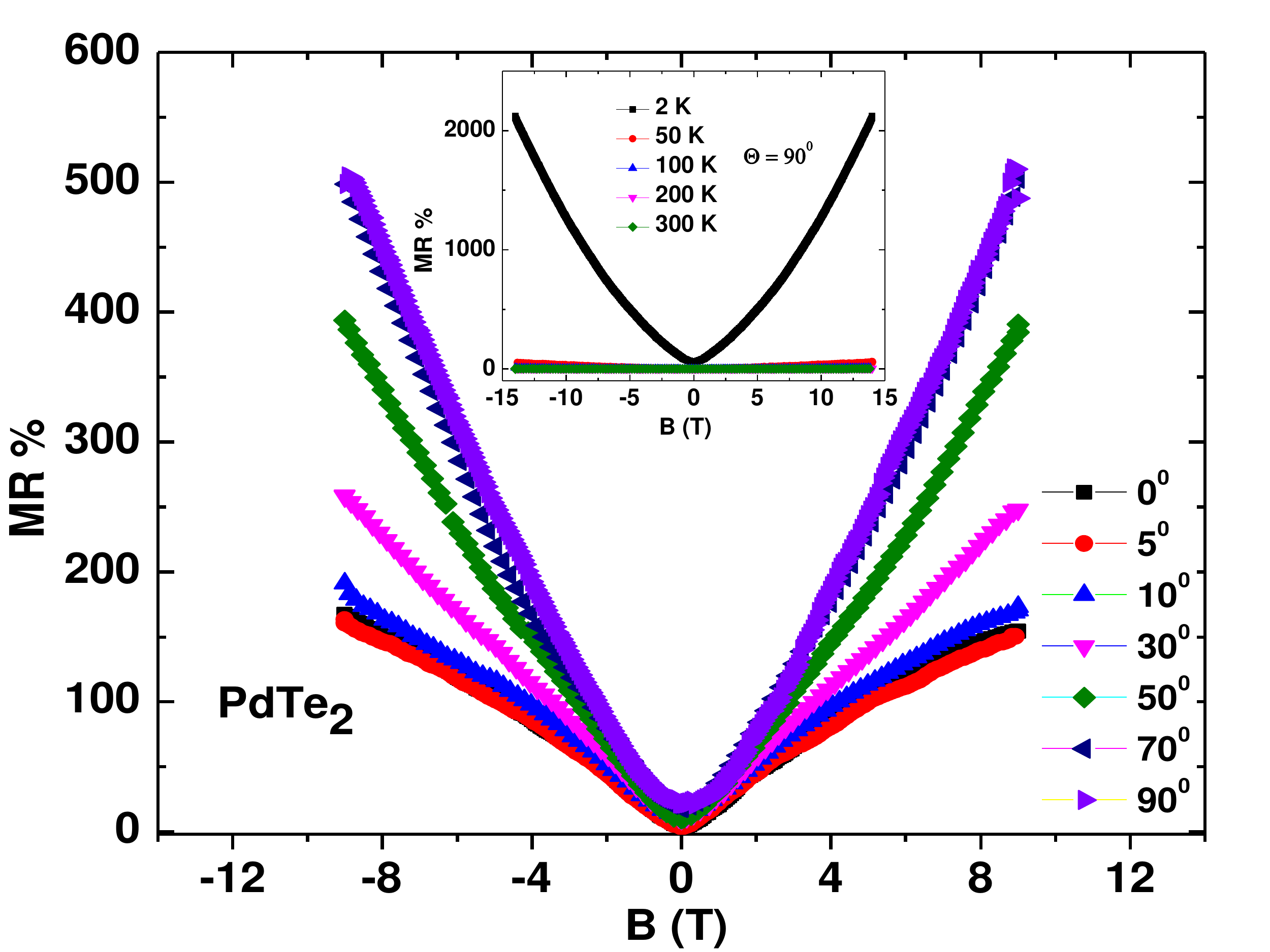}
 	\caption{ Magnetoresistance (MR) for a PdTe$_2$ single crystal as a function of the magnetic field $B\leq 9$~T applied at various angles to the direction of the current $I$ which was always applied within the $ab$-plane.  Inset shows the MR measured at various $T$ with $B \perp I$ for $B \leq 14$~T\@.} 
\label{Fig-MR-Pd}
 \end{figure}

In order to compare the Dirac nature of the bands in PtTe$_2$ with that in PdTe$_2$, we have performed similar magneto-transport measurements and analysis on PdTe$_2$ single crystals. Figure~\ref{Fig-MR-Pd} shows the magneto resistance (MR) data on a PdTe$_2$ crystal for which $B = 0$ electrical resistivity was shown in the lower inset of Fig.~\ref{Fig-res}.  The main panel in Fig.~\ref{Fig-MR-Pd} shows the magnetic field dependence of the MR measured at $T = 1.8$~K as a function of the angle between $B \leq 9$~T and the direction of the electrical current $I$, which was always applied in the same direction within the $ab$-plane of the crystal.  The MR for all angles increases monotonically with $B$.  The magnitude of MR reaches large values of about $600\%$ for $B \perp I$.  As for PtTe$_2$, no negative contribution to the MR was observed for $B \parallel I$ indicating the absence of the Chiral anomaly.  The MR as a function of $B \perp I$ at various temperatures $T$ is shown in the inset of Figure~\ref{Fig-MR-Pt} and shows that the magnitude of MR has a very strong $T$ dependence dropping drastically as one increases $T$ from $1.8$~K\@.

 \begin{figure*}[t]
	\centering
	\includegraphics[width=1\textwidth]{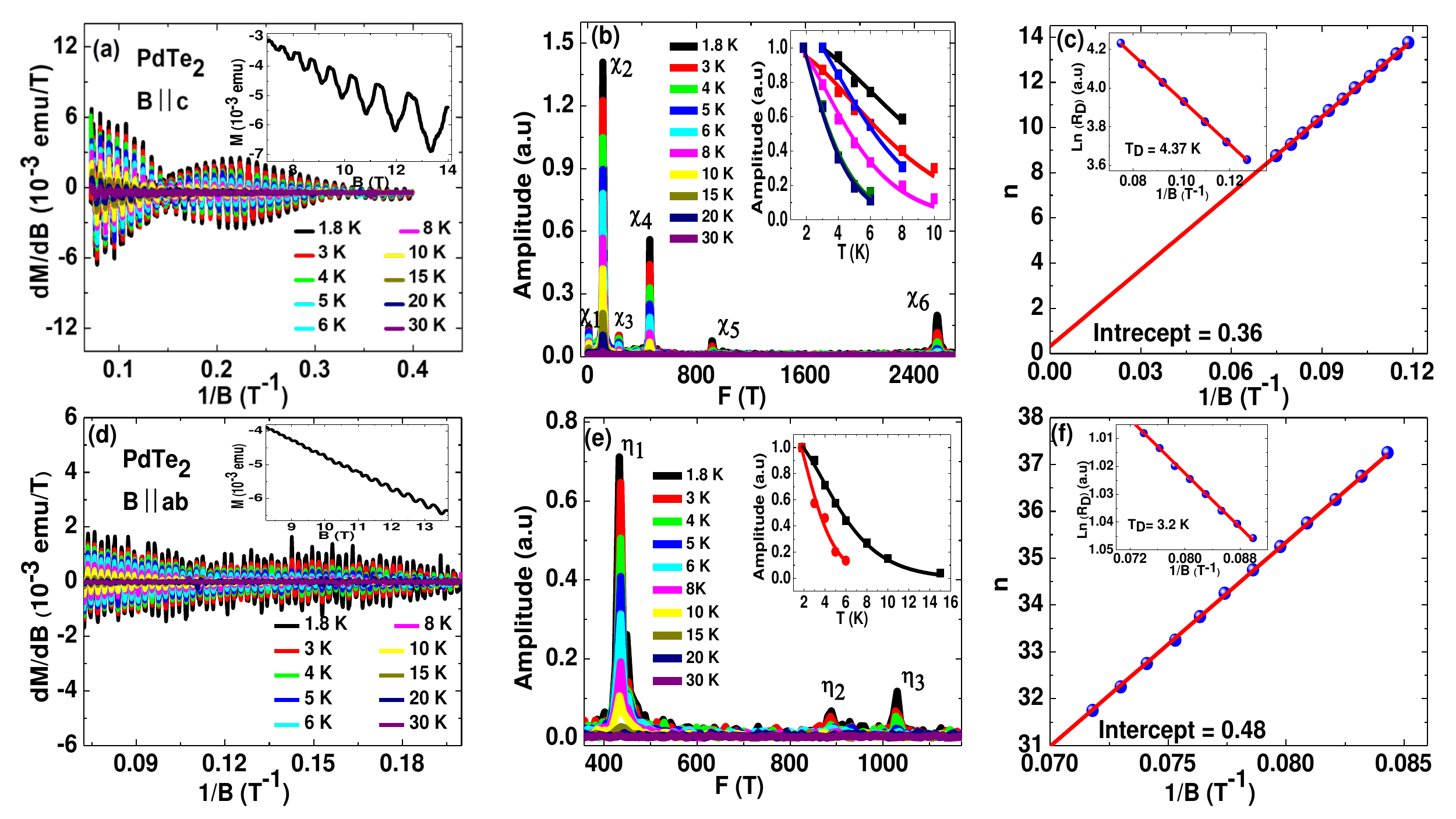}
	\caption{(a and d) Similar set of magnetization data for PdTe$_2$. Magnetization oscillation data for PdTe$_2$ in the two directions B//c and B//ab, at various temperatures.Inset shows row magnetization data at 2K. (b and e) shows the temperature dependence of FFT spectra and fitting of the temperature dependent amplitude is shown in the inset for the two cases. (c and f) The landau level fan diagrams for the two cases. Inset shows the Dingle fitting. } 
	\label{Fig-dM-Pd}
\end{figure*}

\begin{table}	
\caption{Fermi surface parameters for PdTe$_2$ obtained from the dHvA frequencies shown in Figs.~\ref{Fig-dM}~(b) and (e)}.
		\scalebox{0.84}{
			\begin{tabular}{|c|c|c|c|c|c|} 
				\hline
				Compound &	F (T) & $A_f (\AA^{-2})(10^{-2})$ & $K_f (\AA^{-1}) (10^{-2})$ & $m^*/m$ & $V_f (m/s) (10^{5})$\\ [0.ex] 
				\hline
				$PdTe_2, 
				B \parallel c$&	9.13 & $0.087$ &  $1.7$& 0.14& $1.4$\\ 
				\hline
				&112.7 & $1.1$  &$5.84$&0.18& $3.7$\\
				\hline
				&228.7 &$2.2$ & $8.4$  & 0.22& $4.4$ \\
				\hline
				&456.9 & $4.37$  &$11.8$& 0.27& $5$ \\
				\hline
				&913.9 & $8.75$  &$16.7$& 0.41 & $4.7$\\
				\hline
				&2568 & $24.6$  &$28$& 0.43 & $7.5$ \\
				\hline
				$PdTe_2, 
				B \perp c$&	435.8 & $4.17$ &  $6.5$& 0.26& $2.9$\\ 
				\hline
				&889 & $8.5$  &$16.5$&$-$& $-$\\
				\hline
				&1035.5 & $9.8$  &$17.7$&0.46& $4.4$\\ \hline
				
			\end{tabular}}
			\label{Table Pd1}
			\end{table}			
			
			\begin{table}	
				\caption{Parameters obtained from a Dingle fitting to the dHvA data shown in Fig.~\ref{Fig-dM-Pd}}.			
				\begin{tabular}{|c|c|c|c|c|} 
					\hline
					Compound &	$T_D (K)$ & $\tau (10^{-13} s)$ &  l (nm) & $\mu (cm^2/V-s)$\\ [0.ex] 
					\hline
					$PdTe_2, B \parallel c$&	4.37 & $2.8$ &  103.6& 2735\\ 
					\hline
					$PdTe_2, 	B \perp c$&	3.2 & $3.8$ &  110& 2570\\ 
					\hline					
				\end{tabular}
				\label{Table Pd2}
				\end{table}

Figure~\ref{Fig-dM-Pd} shows the magnetization versus $B$ data for PdTe$_2$ for both $B\parallel c$ and $B \parallel ab$ directions, measured at different temperatures.  Figures~\ref{Fig-dM-Pd}~(a) and (d) show the pronounced dHvA oscillations for both directions of magnetic field. It should be noted that the amplitude of the quantum oscillations in PdTe$_2$ is relatively larger than in PtTe$_2$ single crystals and persist up to a temperature more than $30$~K\@. The presence of the multiple frequencies in the two field orientations indicates an even more complex bulk band structure for PdTe$_2$ consisting of multiple Fermi pockets at the Fermi level. The FFT of the oscillations resolves the presence of $6$ main frequencies  for $B \parallel c$ which we label as $\chi_1 (9.13~$T), $\chi_2 (112.7~$T), $\chi_3 (231.5~$T), $ \chi_4 (456.9~$T), $\chi_5 (913.9~$T), and $\chi_6 (2568~$T), and $3$ major frequencies for $B \parallel ab$ which we label as $\eta_1(435.8~$T), $\eta_2(889~$T), and $\eta_3(1030.5~$T). respectively.  The calculated effective masses for the different frequencies for $B\parallel c$ are listed in Table~\ref{Table Pd1} and lie in the range $0.14$--$0.43$. The value of the cross sectional area calculated using the Onsager relation for the lowest effective mass Fermi orbit $\chi_1$ comes out to be $0.87\times10^{-3}$, which is very small compared to the area for other orbits observed for $B\parallel c$. The value of the lowest effective mass for the $B\parallel ab$ direction is found to be $0.26$ which is higher than estimated for $B\parallel c$ direction.  Despite the difference in the effective masses in the two directions, the mobility calculated from a Dingle fitting for the two directions are very similar as shown in Table~\ref{Table Pd2}, in sharp contrast to the case of PtTe$_2$.  These results are consistent with a recent magnetization and ARPES study on the PdTe$_2$ single crystals\cite{16}.  

The Landau level fan diagrams for the two directions are given in Figs.~\ref{Fig-dM-Pd}~(c) and (f).  The intercepts are $0.36(2)$ and $0.48(2)$ for $B\parallel c$ and $B \parallel ab$, respectively.  The corresponding values of the Berry phase are $\phi_B = 0.97(4)\pi$ [or $0.47(4)\pi$] and $1.21(4)\pi$ [or $0.71(4)$] for $B\parallel c$ and $B \parallel ab$.  The value $\phi_B$ for $B \parallel c$ is very close to $\pi$. The Berry phase for $B\parallel ab$ deviates from the value $\pi$. We note that the Fermi level in PdTe$_2$, like in PtTe$_2$, is too far above from the 3D Dirac point which is observed in the ARPES measurements\cite{16,19,20,17}.  
 
\subsection{Conclusion:}
We have presented a detailed magneto-transport study on single crystals of the di-Tellurides PtTe$_2$ and PdTe$_2$ with an emphasis on trying to ascertain the possible topological nature of the bands contributing to the transport. 
Prominent dHvA quantum oscillations are observed for both materials in both directions of applied magnetic fields.  From an analysis of the magnetization data on the two materials, it is found that the Fermi surface of both systems is highly anisotropic in nature which is evident from the different number of the oscillation frequencies in the two field directions.  Additionally for PtTe$_2$ a very large difference in the value of the mobility (more than three times) in the two crystal orientations is observed and would be consistent with expectations of highly anisotropic transport resulting from the tilted nature of the Dirac cone in the  PtTe$_2$.  The Berry phase for PtTe$_2$ is close to $\pi$ for $B \parallel c$ while it deviates from $\pi$ for $B \parallel ab$.  We speculate that this could be due to a combination of the Fermi level being away from the Dirac point and the presence of other topologically trivial bands at the Fermi level.  This anisotropy is almost absent in case of  PdTe$_2$ as the calculated mobility in the two crystal orientations is of the same magnitude. The bands in  PdTe$_2$ are indeed three dimensional Dirac bands characterized by the Berry phase close to $\pi$ in both in plane and out of plane crystal orientations.
Our results therefore suggest that PtTe$_2$ like PdTe$_2$, could also be a Type-II Dirac semi-metal.

\begin{figure}[h!]
 	\includegraphics[width=0.52\textwidth]{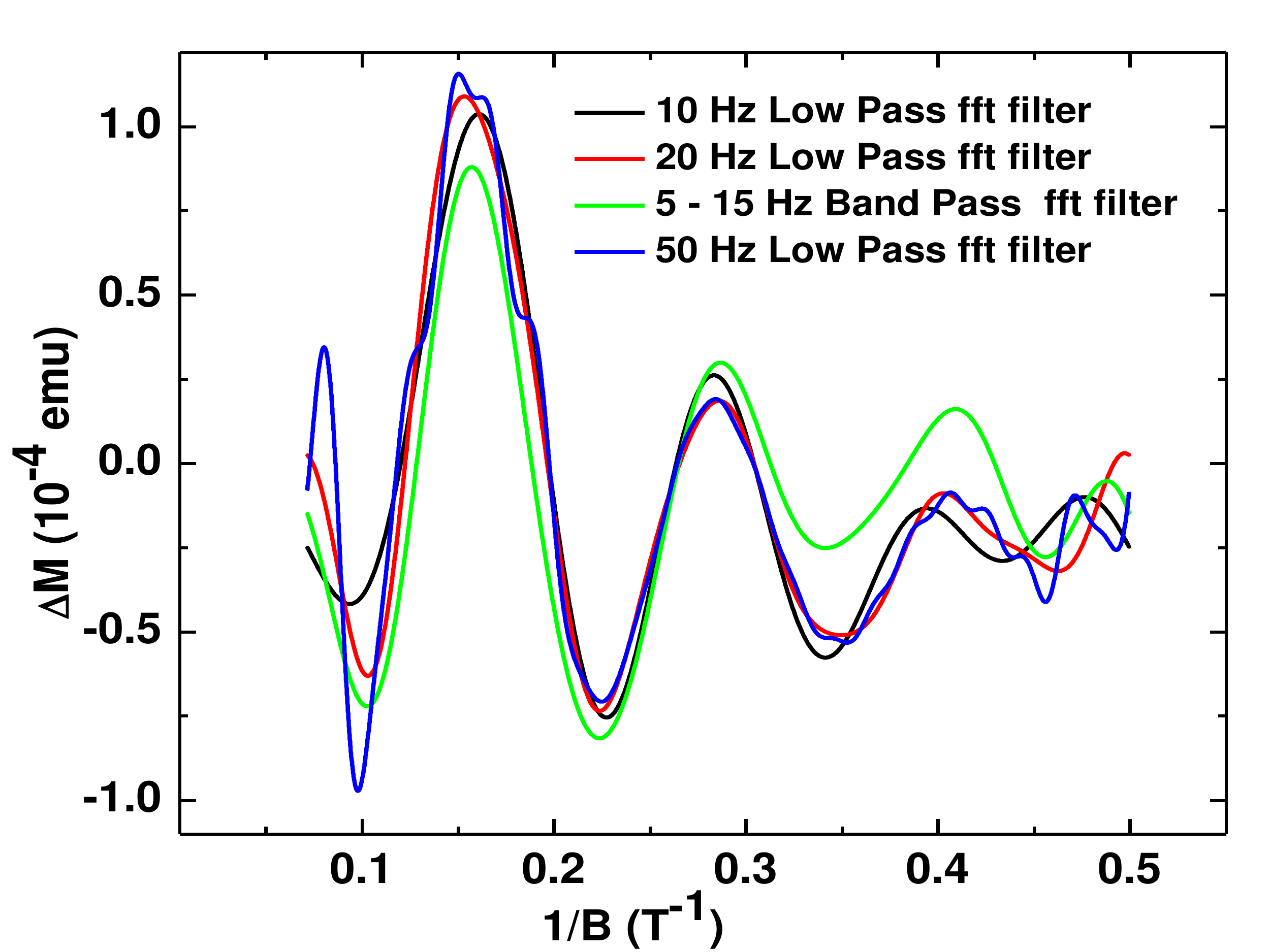}
 	\caption{ Oscillatory signal obtained after applying various low-pass filters to the dHvA data for PtTe$_2$ in the $B \parallel c$ direction.  The oscillations clearly depend on what filter is applied.} 
\label{Fig-filter}
 \end{figure}

\noindent
\emph{Note added:} While this manuscript was under preparation another quantum oscillation study on $A$Te$_2$ ($A =$ Pt, Pd) has appeared on the arxiv \cite{Zheng2018}.  This work also reports low carrier effective masses and high mobilities for the two materials.  Their Landau level fan diagram analysis however, is different.  In order to obtain the contribution from an individual orbit out of the multi-frequency quantum oscillation data, they have used a low-pass filter of $50$~T to get the contribution from the lowest frequency orbit $8$~T for PdTe$_2$ in the $B \parallel c$ direction.  From the Landau level fan diagram obtained from this they conclude that the Berry phase for this orbit is different from $\pi$.  \\
We point out that by applying a low-pass filter, one can not get a unique Landau level diagram as the location of the maxima and minima in the oscillation data obtained after applying the filter depend on what kind of filter has been applied.  In Fig.~\ref{Fig-filter} we show the result of applying different low pass filters to the dHvA data for PtTe$_2$ for $B \parallel c$.  It is clear that different filters give oscillations with slightly different locations of the maxima and minima, which are used to construct the Landau level fan diagram.  This suggests that the Landau fan diagram and any numbers obtained from an analysis of the same will strongly depend on the filter applied.    

\end{document}